\documentclass[11pt]{article}
\usepackage[utf8]{inputenc}
\usepackage{amsmath,amssymb}
\usepackage{graphicx}
\usepackage{hyperref}
\usepackage{cite}
\usepackage{geometry}

\usepackage{color}
\definecolor{myR}{RGB}{229, 0, 33}

\newcommand{\FIGS}[1]{figsmedium/#1}

\title{On Quad Mesh Extraction From Messy Grid Preserving Maps}
\author{Nicolas Ray\\ Pixel, Université de Lorraine, CNRS, Inria, LORIA, 54000, Nancy, France}
\date{}

\begin{document}

\maketitle

\section*{Abstract}

Extracting a quad mesh from a grid preserving map is straightforward in theory, but typical inputs are not exactly grid preserving maps. Previous works can manage minor deviations from grid preserving maps, but without a clear specification of what is acceptable. This work clarifies how typical inputs differ from a grid preserving map, and shows how the differences with a grid preserving map can be reflected by a sequence of operations acting on a discrete structure. It opens research opportunities for the design of a robust quad extraction algorithm.

\section{Introduction}

Quad mesh generation is a challenging task, especially when aiming for a block-structured layout. In such cases, a possible approach is to view the quad mesh as a deformed grid that covers the target surface. Quad meshing then becomes a problem of defining this deformation, which is usually done by computing a map i.e., the inverse of the grid deformation. To handle complex surfaces, discontinuities must be introduced into these maps. Provided these discontinuities are \emph{grid preserving} (GP) \cite{Lyon2019}, they will not be noticeable once the 2D grid is projected onto the surface.

\begin{figure}[ht]
    \centering
    \includegraphics[width=0.7\linewidth]{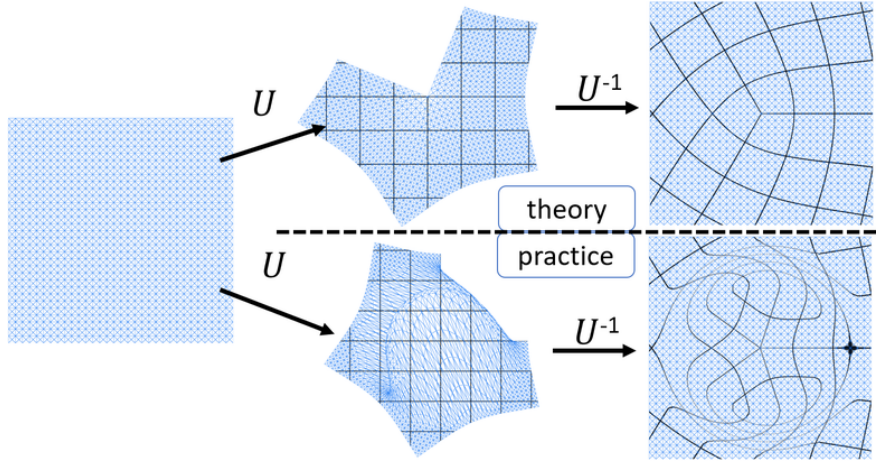}
    \caption{Given a function $U$ that maps the geometric domain (left) to a 2D space (middle column), one can use the invert function $U^{-1}$ to project the unit grid on the surface. If the map is injective (upper row) the result is a deformed grid equivalent to a quad mesh, but it is less obvious for lower quality maps (lower row).}
    \label{fig:theory}
\end{figure}

Theoretically, if we can compute a Foldover-Free GP map and overlay a regular grid on the map, its pre-image will segment the surface into charts with four boundaries --- forming a valid quad mesh as depicted in figure \ref{fig:theory}. However, in practice, the foldover-free property is difficult to ensure.

In the optimistic case of having a foldover free map as input, extracting a quad mesh from a GP map is still non-trivial due to numerical issues: floating-point inaccuracies and the complexity of tracing iso-lines that intersect with the mesh primitives (vertices, edges, triangles). Robust extraction requires exact predicates and the careful handling of all intersection configurations. The state of the art of quad mesh extration \cite{QEX,Hexex} address these issues efficiently, and propose solutions to handle some cases of foldovers. 

As map quality deteriorates, the hope of extracting a quad mesh with good geometry quickly vanishes. However, some structural (combinatorial) properties of the mesh may still be preserved. As shown in \cite{MIQ}, GP maps are defined by real-valued variables, but are subject to grid-preserving constraints that depend on integer variables. Intuitively: integer variables define the \textbf{combinatorial structure} of the quad mesh, whereas real variables define its \textbf{geometry}.

This leads to the hypothesis that it should be possible to recover a quad mesh with the same combinatorics whatever the map distortion due to real variables could be. We study the possibility to extract such a quad mesh from a messy GP map.  

\subsection*{Related works}

Quad extraction was first introduced as a post-processing step of grid preserving maps algorithm \cite{quadcover,MIQ}. A few years later, \cite{QEX} observes some failure cases in quad extraction and solves those related to numerical approximations, and some of those due to foldovers in the map. The idea is to trace iso curves of integer value of map coordinates onto the surface, and to extract a quad mesh from this network of curves. The method was improved and extended to volumes in \cite{Hexex,Kohler:2025:hexhex}. Both have author's code available that we use for comparisons. 

The litterature for quad mesh extraction is rather thin because it resolves issues that could be dealed with somewhere else in the pipeline of quad mesh generation:
\begin{itemize}
    \item Quad mesh extraction originally assumes that the input is valid. The typical problem is that maps are not likelly to be injective (without foldovers) if the map generation algorithm does not focus on this specific constraint. There exists a large litterature to enforce it, for examples \cite{Fu2017,Garanzha2021}. Another problem tackled in \cite{Garanzha2024} is that extra singularities may appear on vertices of the triangle mesh.
    
    For maps with grid-preserving discontinuities, there is much less work about preserving injectivity. A solution was proposed by \cite{IGM2013}, but it was not the main contribution of this work, and the robustness against hard cases was not evaluated. Finally, the problem without discontinuities is very difficult, adding seamless maps constraints makes it harder, and grid preserving is even worse, especially if a coarse mesh is expected.

    Actual solutions are not $100\%$ robust, are difficult to implement and tune, and are quite slow to run. Moreover, injective maps may not always exist under the discretization constraint of being linear on each triangle.

    \item Another solution is to completely skip the quad mesh extraction step of the pipeline. Once integer degrees of freedom $k_{ij},T_{ij}$ (see \S\ref{sec:pb_settings}) of the map are found, \cite{Lyon2019} prefers to skip the computation of a new map and to directly work on the T-mesh decomposition used to compute these integer degrees of freedom. It requires re-embedding a T-mesh already living in the original surface, but it is nevertheless considered as an acceptable atlernative to compute a map and extract a quad mesh out of it.
\end{itemize}

Being able to generate a quad mesh out of low quality maps would allow to increase the success rate of the pipeline, despite using simple/naive map generation algorithms. Moreover, it provides insight for 3D extension where \cite{Hexex} already made it possible to generate meshes from maps obained from invalid frame fields.

\subsection*{Contributions}

This work allows to better understand the properties one can expect when trying to extracted a quad mesh from a messy input map.  

\begin{itemize}
    \item we analyse the conditions for uniqueness of quad mesh that could be extracted from a given map \S \ref{sec:pb_settings}
    \item we introduce the \textbf{dual segmentation} \S \ref{sec:dual_segmentation}, that allows to extract a quad mesh from a valid GP map, and that reflect map deformations by two combinatorial operations. It opens research opportunities to design a robust algorithm by inverting these operations.
    %\item we propose a new algorithm based on inverting these operations \S \ref{sec:algorithm}. It does not solve all possible cases, but covers more of them than previous works. 
\end{itemize}

The remainder of the paper is organized as follows: we first clarify what is expected as input \S \ref{sec:pb_settings} and how it affects the uniqueness of extracted quad meshes, then we introduce the dual segmentation and show how a set of canonical operations reflects modifications of the map \S \ref{sec:dual_segmentation}. In appendix \ref{sec:algorithm}, we show that the inversion of these operations can produce a simple and competitive algorithm, despite being limited to a subset of configurations.

\section{Problem Settings}
\label{sec:pb_settings}

Quad mesh extraction is well defined with perfect inputs. For more practical cases, we need to define how boken could be the inputs, and what is the quality of ouput we could expect from those inputs.

\subsection{Inputs}

\begin{figure}[ht]
    \centering
    \includegraphics[width=0.7\linewidth]{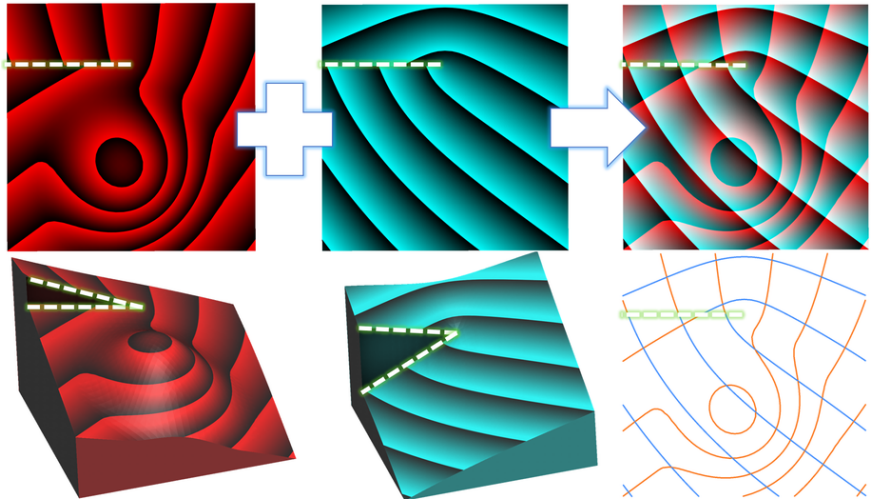}
    \caption{{\bf Map visualization}. Here, the surface is a square, and map coordinates are two functions defined over the square (left and middle rows). Each function can be plot in 3D with colors to better visualize the iso values (lower), but are often depicted using only the colors (upper). It is then possible to combine both functions by merging their colors (upper--right) to represent the map on a single image. Iso integer values of both field (lower--right) represents the segmentation of the domain. Notice that the GP discontinuity is visible as a cut in the plots, and produces a color swap in right images.}
    \label{fig:render}
\end{figure}

Our input is a map produced by an algorithm that correctly fixed the integer degrees of freedom of the optimization problem described in~\cite{MIQ}. Let $F : S \rightarrow \mathbb{R}^2$ be a map defined over a triangulated surface $S$ (illustrated as in Figure \ref{fig:render}). It may satisfy the following properties:

\begin{itemize}
    \item \textbf{Seamless:} The transition function on the edge between triangles $i$ and $j$ satisfies $F_i = R^{k_{ij}} F_j + T_{ij}$, where $R$ is a rotation of $\pi/2$, $k_{ij} \in \mathbb{Z}$, and $T_{ij} \in \mathbb{R}^2$.
    \item \textbf{Grid Preserving (GP):} seamless map where $T_{ij} \in \mathbb{Z}^2$.
    \item \textbf{Foldover-Free (det+):} $\det (\nabla F) > 0$ everywhere. The Jacobien of the map is positive everywhere means that there is no foldover in the map.
    \item \textbf{Singularity on Grid (SoG):} Singularities are mapped to integer grid points: $F(v) \in \mathbb{Z}^2$ for singular vertices $v$.
\end{itemize}

The \textbf{det+} property ensures that inside each triangle the projection of the 2D unit grid onto the surface is a deformed grid. The \textbf{GP} property ensures the continuitiy of the grid across edges and, coupled with \textbf{det+}, ensures that the grid has no pleat located on edges. The \textbf{SoG} property ensures that singularties are located on vertices of the grid. Finally, vertices with valence $\neq 4$ are allowed, but polygons with something else than 4 edges are not (it would not be a quad). A unique quad mesh can be computed \cite{QEX} from a map with all these properties (\textbf{GP}, \textbf{det+}, and \textbf{SoG}).

Unfortunately, typical inputs do not have all these properties. They are always \textbf{GP} because it is enforced via frame fields and integer quantization (determining $k_{ij}$ and $T_{ij}$), even if the quantization is not valid as defined in \cite{QGP}. However, \textbf{det+} is more difficult to guarantee as it is expressed as a quadratic inequality constraint in the numerical optimization of the map.

Surprising, the case of \textbf{SoG} is even more difficult to ensure because the map may have singularities that were never prescribed for the map optimization. Actually, most singularities have their map coordinates already enforced by the transition functions ($k_{ij}$ and $T_{ij}$).However, it is not sufficient to avoid singularities to appear at integer $+\frac{1}{2}$ coordinates, and for singularities of integer index like -1 for a saddle. For those cases, it is necessary to explicitly enforce map coordinates of known singularities in the optimization problem. A more fundamental difficulty for \textbf{SoG} comes from singularities that did not exist in the frame field, but rather appeared during the numerical optimization that produced the map (see \S \ref{sec:maponlysingu}).

\subsection{Output}

Given a \textbf{GP} map $U$, we assume that there exists at least one map $M$ satisfying \textbf{(GP, det+, SoG)} with the same $k_{ij}$ and $T_{ij}$ as $U$. If so, we can define a \textbf{noise map} $N = U - M$, such that $N$ is \textbf{GP} (same $k_{ij}$ than in $M$, and $T_{ij} = 0$), and $N$ vanishes on the singularities of the frame field.

The \textbf{expected output} is the quad mesh that would be extracted from the map $M$. Starting from our a map $U = M + N$ ($M$ and $N$ being unknown), our objective is to remove the perturbations due to $N$.

In this problem, the expected output is not unique because its geometry depends on the map $M$ that is never unique, and its combinatorics may change if the combinatorics of $M$ allows some singularities to move (see \S \ref{sec:maponlysingu}). On the other hand, a valid $M$ may not exist due to incoherent sets of integer variables $k_{ij}$ and $T_{ij}$, respectively produced by the frame field and the quantization steps of the map generation algorithm.

\subsection{Map-Specific Singularities}
\label{sec:maponlysingu}

A \textbf{det+} map will generate a deformed, regular grid on each triangle, and the \textbf{GP} property ensures that the grid is continuous across edges. Therefore the grid is regular everywhere, but on vertices. Given a vertex $v$, we can characterize its regularity by its index defined as $2\pi$ minus the sum of angles of adjacent triangle corners mesured in the map, over $2\pi$. The index of $v$ is obviously related to the transition functions of the edges having $v$ as an extremity, and all indices are necessary multiple of $\frac{1}{4}$. However, the set of $k_{ij}$ defines only the fractionnal part of the index, not its integer part. This is due to $R^{k_{ij}}$ being a matrix of rotation of $k_{ij} \pi /2 \mod 2 \pi$, where the modulus makes $k_{ij} \equiv k_{ij} \mod 4$ in the constraints.

Let's consider a case where $k_{ij}=0, \forall i,j$: the vertex $v$ can have any non positive integer index (positivity is incompatible with \textbf{det+}). For example, the map given by the complex function $\exp{2 \log{z}}$ is null at 0, has no discontinuities ($k_{ij}=0 \forall i,j$) and an index $1$ at $0$, whereas the identity function have the same properties, but and index $0$ at $0$. This situation commonly occurs with free boundary maps and null fractionnal part, as exposed in \cite{Garanzha2024}.

Given that our input is only assumed to be \textbf{GP}, the integer part of indices is not specified by our input, leading to many possible combinatorics. However, typical inputs are not an isolated singularity inside a domain without boundary conditions. The context has to be taken into account in order to analyse how such singularities may appear during the map optimization process. Using our map decomposition $U=N+M$, the question is: how could $N$ generate indices in $U$ that are not those of $M$ ?   

\begin{figure}[ht]
    \centering
    \includegraphics[width=0.7\linewidth]{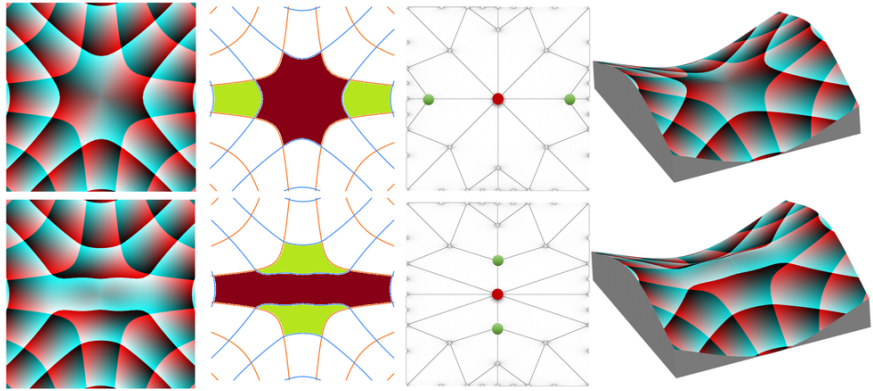}
    \caption{A (det+) map can have a saddle in each of its map coordinate (upper line), leading to a index -1 singularity. It is illustrated by (from left to right): merged color of each coordinate, segmentation by the iso integer curves, quad mesh dual to the segmentation, and a 3D plot of the map coordinate represented by the blue color. Locally lifting one coordinate (blue field and iso-values) changes the connectivity of the retro-projected grid (lower line) and, therefore, the extracted quad mesh.}
    \label{fig:lift}
\end{figure}

In fact, Poincare Hopf theorem would not allow a new index to created. What $N$ can do to perturbate the indices of $M$ is an operation that we will name as saddle lift and that basically reconnects two iso-curves. The simplest case is illustrated in Figure \ref{fig:lift} where the map has a saddle in each map coordinate, and we lift it in one of the coordinates. This operation affects the combinatorics of the quad mesh to be extracted, but does not change the indices of singularities... to observe it, we need to combine saddle lifts with singularities of other indices than $-1$ (see Figure \ref{fig:lift}).

This later configuration occurs very often: especially with indices $-1/2$ or $-1/4$ that can easily be converted into combination of indices $-1$ and $1/2$ (resp. $3/4$) singularities to reduce the map distortion. Unfortunately, these indices are not even valid for producing a quad mesh (it generates valence 1 and 2 vertices).

\begin{figure}[ht]
    \centering
    \includegraphics[width=0.7\linewidth]{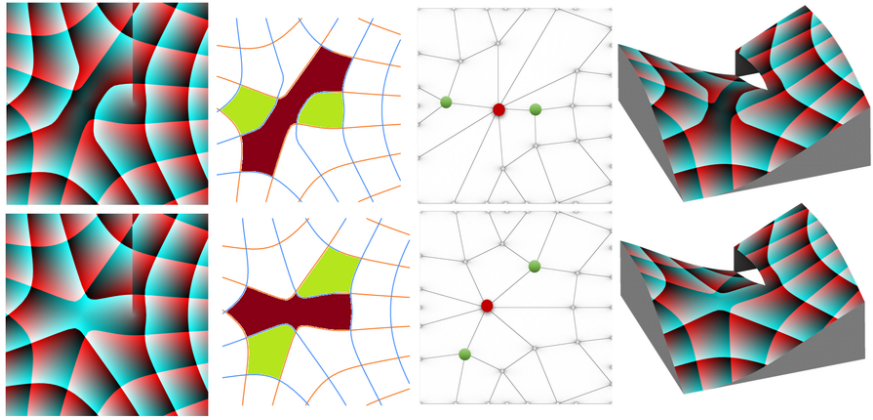}
    \caption{A saddle lift as in Figure~\ref{fig:lift} may interact with existing singularities, by changing their indices.}
    \label{fig:lift2}
\end{figure}

\textbf{Familly of maps:} Given a global map, every possible saddle lift will produce a quad mesh with another combinatorial structure. However, we conjecture that, on a closed surface, there is a unique combinatorial structure for a given set of indices of singularities. For open surfaces, a similar condition must be respected on each boundary.

\subsection*{Notations:} A map is a function defined on the input surface that gives the 2D map coordinates. It can be separated into two scalar fields $u$ and $v$ of the surface.

{\bf Iso:} A iso-u (resp iso-v) is an iso-value of the scalar field $u$ (resp. $v$), taken with an integer value.

{\bf Index and valence:}  In our context, the index of singularities can be interpreted as the expected valence of the quad mesh vertices: $Valence(v) = 4 - 4 \times Index(v)$.

\section{Dual Segmentation}
\label{sec:dual_segmentation}

Given a (det+, GP,SoG) map $M$, the surface can be segmented into charts that correspond to the projection by $M^{-1}$ of the unit grid: the 2D grid with quadrangular cells of coordinates (i,j),(i+1,j),(i+1,j+1) and (i,j+1). The approximation of each chart by a bilinear quad is the quad mesh to be extracted.

The classical way to extract a quad mesh (\textbf{Primal segmentation}) is a direct implementation of this definition. Chart edges are computed on the surface as iso curves of both map coordinates, extracted at integer values. This approach gives precise vertex positions and is straightforward to convert to quads.

We introduce the \textbf{Dual segmentation}, where the map is translated by $(1/2,1/2)$ before extracting the iso-curves. The resulting segmentation corresponds to the \emph{dual} of the quad mesh i.e. charts correspond to quad vertices and vertices of this segmentation correspond to quads. Its combinatorial structure is equivalent to the one given by the primal, but the location of quad vertices is not directly available. Working on the dual is also very common in overlay grid methods (2D and 3D) to better capture the boundary \cite{dualcontouring2002,Gao2019}.

\begin{figure}[ht]
    \centering
    \includegraphics[width=0.5\linewidth]{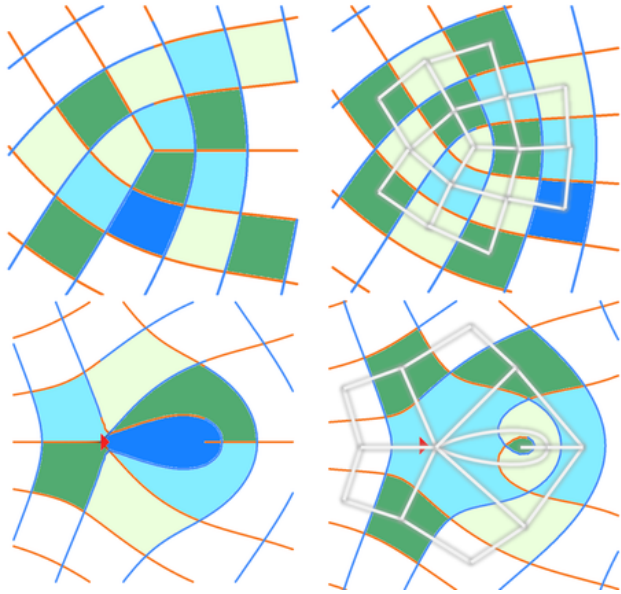}
    \caption{Primal (left) versus dual (right) segmentation. The primal is directly the quad mesh for simple cases (upper row). When the map geometry nearby singularities is degenerated, reverted or numerically instable, the structure of the primal is less straightforward to extract. The dual is always well defined because its frontiers are extracted far away from these regions.}
    \label{fig:dual}
\end{figure}

\begin{figure}[ht]
    \centering
    \includegraphics[width=\linewidth]{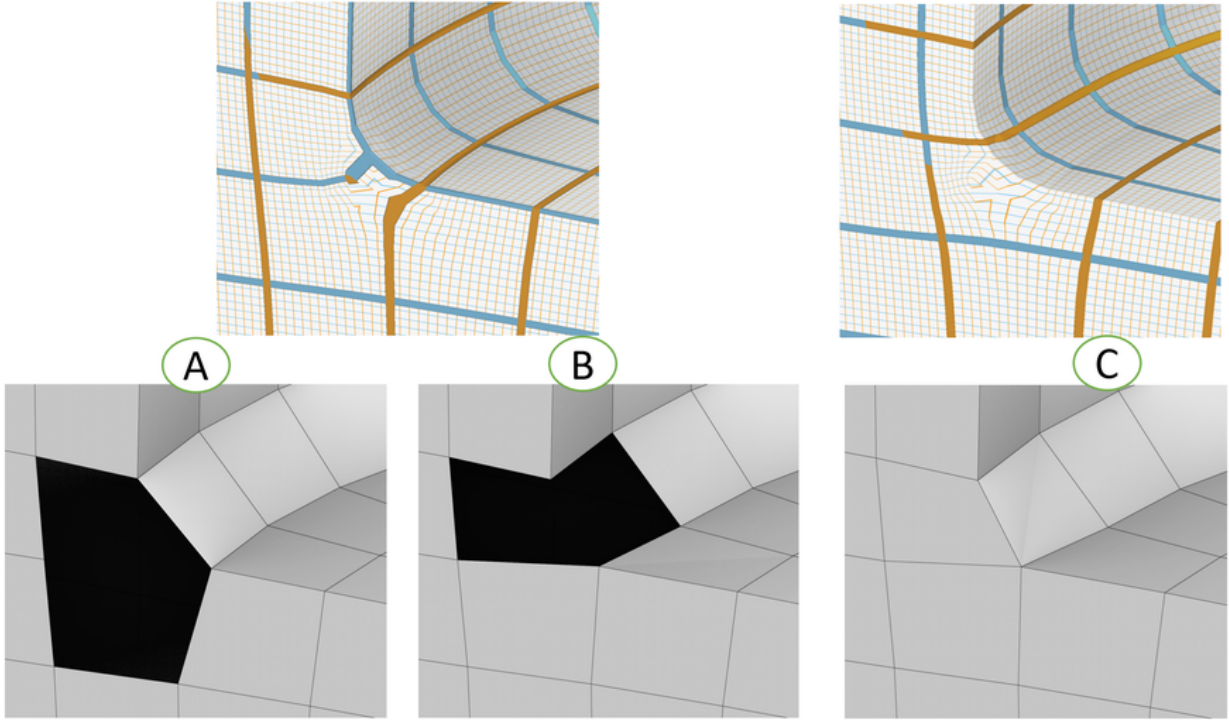}
    \caption{Foldovers often occurs with complex configurations close to singularities. As they are located on primal chart boundaries, the primal (left) is likely to be more complex than the dual (right). Qex (A) and Hexex (B) failed to get a correct mesh from the primal segmentation, whereas our algorithm based on the dual succeed.}
    \label{fig:primaldual}
\end{figure}

Benefits of the dual segmentation come from the regularity of its structure: iso curves are always far from locations where integer constraints are acting i.e. singulary vertices, boundaries or feature curves. As a consequence, there is no dandling edges or noisy regions as illustrated on Figures \ref{fig:dual} and \ref{fig:primaldual}. Moreover, vertices of this segmentation are necessary the intersection of two iso-curves, and directly correspond to quads on the final mesh when there is no foldovers. 

For these reasons, we use the dual segmentation as a combinatorial structure that abstracts the map $U$ and allows the impact of the noise map $N$ to be analyzed.

\subsection{Dual segmentation of maps corrupted by noise}

Our input maps can be decomposed as $U=M+N$ where we assume that the noise map $N$ is a sum of smooth, local perturbations, each being defined over a topological disk free of singularity. This allows the effect of $N$ to be interpreted as a sequence of local modifications of the segmentation that would correspond to $M$.

\subsubsection*{Local Noise on Disks}

By construction, $N = 0$ on the singularities of the frame field. Thus, $N$ can be split into local perturbations, each supported on a disk that does not contain a singularity with a non integer index.

Each local perturbation been supported on a topological disk, we can fix the map coordinates on one triangle and greedily change its neigborgs map coordinates to remove disconinuties ($k_{ij}=0, T_{ij}=0$) within the disk. By doing so, the maps limited to such disks are simply defined by two continuous scalar fields $u$ and $v$ that represents the map coordinates.

\subsubsection*{Impact of the noise on iso-curves}

\begin{figure}[ht]
    \centering
    \includegraphics[width=0.99\linewidth]{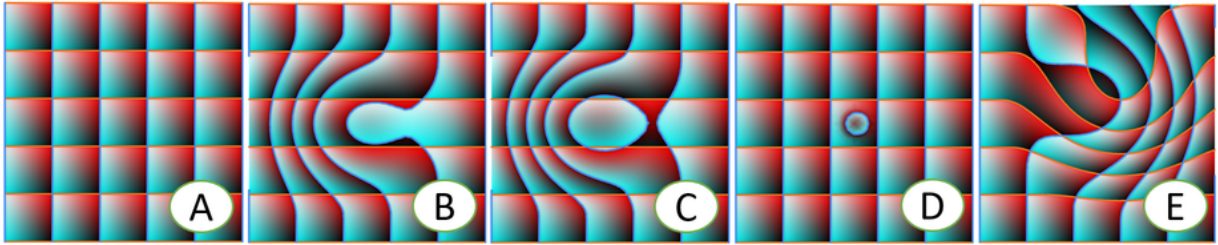}
    \caption{The identity map (A) can be corrupted by noise that: just move the geometry of boundaries(Figure \ref{fig:noiseimpact}--B), \textbf{reconnect iso-curves} when the integer part of the value of a saddle changes in one of the fields (Figure \ref{fig:noiseimpact}--B to C), \textbf{create/delete iso-curves} when the integer part of the value of a local extrema changes in one of the fields (Figure \ref{fig:noiseimpact}--D), or create \textbf{u/v interleaving} when u and v curves touch and overlap each other (Figure \ref{fig:noiseimpact}--E).}
    \label{fig:noiseimpact}
\end{figure}

\begin{figure}[ht]
    \centering
    \includegraphics[width=0.4\linewidth]{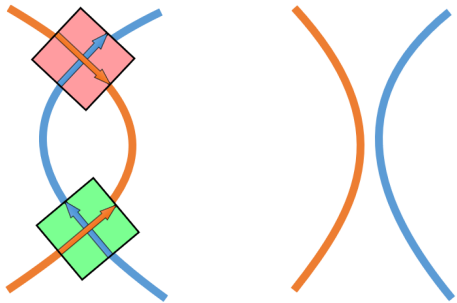}
    \caption{\textbf{OP1 (Untangling):} When a iso-u overlaps an iso-v, one over two intersection is reverted (Left) -- the orientation of the intersection is given by the square color and the relative orientation of the arrows. By moving each iso independently, the overlap can be removed (Right).}
    \label{fig:OP1}
\end{figure}

\begin{figure}[ht]
    \centering
    \includegraphics[width=0.7\linewidth]{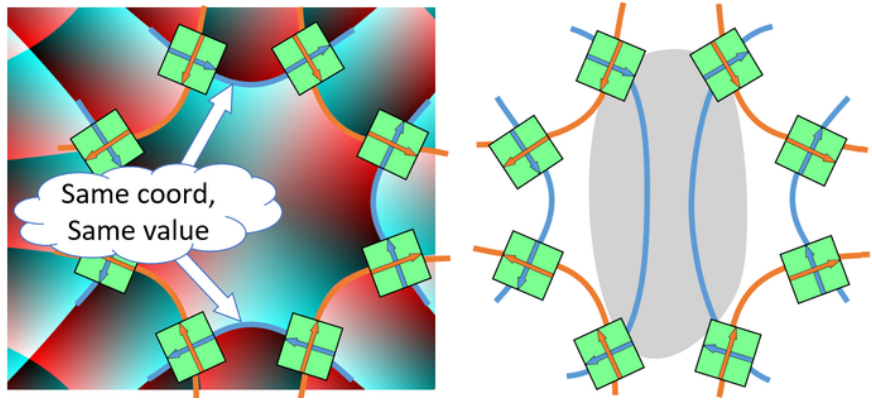}
    \caption{\textbf{OP2 (Saddle lift):} Two iso of the same coordinate, with the same value (Left) can be reconnected differently (Right). }
    \label{fig:OP2}
\end{figure}

Within each disk, charts are bounded by iso-$u$ and iso-$v$ curves. Adding local noise to these fields will impact the segmentation as depicted in Figure \ref{fig:noiseimpact}. These changes lead to topological events in the segmentation that can all be represented by three operations:  

\begin{itemize}
    \item \textbf{OP1 (Untangling):} (Figure \ref{fig:OP1}) When overlapping iso-$u$ and iso-$v$ are split apart, they merge three charts into one. Iso-u and iso-v intersect once in a $\det > 0$ region and once in a $\det < 0$ region. The inverse operation is \textbf{OP1--inv (tangling)}.
    \item \textbf{OP2 (Saddle Lifting):} (Figures \ref{fig:lift}, \ref{fig:lift2}, and \ref{fig:OP2}) Reconnecting iso-curves can merge two charts while splitting another. It reflects a saddle value change in the $u$ or $v$ field.
\end{itemize}

\textbf{Remark:} When an iso-curve appears or disappears inside a chart (Figure \ref{fig:noiseimpact}--D), it generates a region with a multiple boundaries. This is a special case of OP2, applied between two segments of the same iso-curve.

\subsection{How far is a robust algorithm ?}

From the analysis of the previous section, we know that the impact of adding noise onto the dual segmentation is limited to a succession of local (reversible) operations OP1, OP1--inv, and OP2. Therefore, the dual segmentation extracted from $U$ could be tranformed into a segmentation of $M$ by applying a correct sequence of operations. At this point, there are (only) two problems to solve to obtain a robust algorithm: have a representation of the dual segmentation, and find a good sequence of operations.

\paragraph{Dual Segmentation Representation}

Let's consider first the most natural representation of the segmentation: a triangle mesh with a GP map and edges placed on iso-lines of integer values of each map coordinates. This structure can be obtained by imprinting iso-values in the original triangle mesh, even with a messy input GP map. In this structure, segmentation charts and their frontiers are easy to compute, which is sufficient for extracting a quad mesh for (GP, det+, SoG) maps.

Moreover, it is easy to check preconditions for OP2 and OP1--inv in this structure: we select the region of interest as a topological disk without singularities defined by a set of triangles, and move each triangles in the map to remove all discontinuities of the map inside the region. Now it is easy to check if the value and coordinates of iso are the same before applying OP2 (or different coordinates for OP1--inv).

Difficulties happen when OP1, OP1--inv or OP2 has to be applied: these operations are defined on the combinatorial structure of the segmentation, but not on its geometry. Therefore this natural representation does not directly supports our combinatorial operations:
\begin{itemize}
    \item A first solution would be to define how our combinatorial operations act at the geometric level, but a good geometry remains to be defined and applying multiple embeddings on a mesh is a tedious task prone to generate numerical errors.     

    \item A second solution would be an explicit representation of the combinatorial structure that is not as simple as a polygonal mesh because charts may have more than one boundary and a frontier between two charts may not have endpoints. The last challenge is that this structure must be able to check preconditions for OP2 and OP1--inv: it must be possible to remove all map coordinates discontinuities in a topological disk region that does not contain a singularity. What makes it challenging is that regions of interest may contain frontiers that do not belong to the same boundary of a chart. 
\end{itemize}

\paragraph{Choosing the sequence of combinatorial operations} 

The second difficulty is to find a sequence of combinatorial operations that converges to a quad mesh without reverted quad. Moreover, if multiple combinatorial structures are possible for the quad mesh, the algorithm should produce a ``good'' one according to some criteria such as non degenerated quads, close to the input map, or respecting singularity index prescription. 

The algorithm is basically a graph search where nodes are segmentations linked by operations. A brute force search is certainly untractable, hopefully there will be effective pruning strategies to solve the problem: 

\begin{itemize}
    \item OP1--inv can be applied everywhere, but creating more reverted regions is opposite to our objective, and we do not foresee any configuration where applying OP1--inv could be required.

    \item OP1 applies only to a reverted intersection linked with a normal intersection. Moreover, if OP1--inv is never needed, OP1 can be applied as soon as possible. At the end of the process, OP1 will be applied exactly once per reverted intersection.
        
    \item A saddle lift (OP2) requires that a chart boundary have two edge sections with the same iso (coordinate and value). We can always take two sections of the same edge, but it's probably not a good move as it simply adds a single edge boundary to the chart. Eligible sections that belongs to two edges are not very common: it can occur nearby singularities of index $<-\frac{1}{4}$ as explained in \S\ref{sec:maponlysingu}, or on edges that would be merged by applying OP1, or inside multi-boundaries charts to merge them.     
\end{itemize}

If OP1 is applied as soon as possible and OP1--inv is never applied, the choices to determine the sequence are reduced to OP2 that will either merge boundaries of a chart, or move a singularity index produced from a singularity of index $\leq-\frac{1}{4}$.

\section{Discussion}
\label{sec:discussion}

When extracting a quad mesh extraction from a GP map, noone would be satisfied with a degenerated/unexpected combinatorial structure, that is nevertheless a valid solution. Therefore, it will be important to move to a different problem where singularities index are prescribed. Another important question is how would the approach extend to the volumetric case.

\paragraph{Unique combinatorial structure}

Map coordinates are not sufficient to define a unique combinatorial structure because saddle lifts may alter it. Fixing map coordinates at singularites would not be sufficient either because saddle lifts would be able to move singularity indices from a singularity to another. We conjecture that a solution to disambiguate the combinatorial structure is to fix the indices of singularities, as it is usually available in a frame field. 

Explicitly storing singularity indices would be another constraint to the representation of the segmentation: singularities are located inside charts and should be preserved by combinatotrial operations.

\paragraph{What about hexahedral meshes ?}

Obviously, the problem studied here also happens in 3D grid preserving maps for extracting hexahedral meshes. It should be possible to extend the 2D theoric analysis into 3D, where topological events on  iso-u, v, and w surfaces will be considered. However, the representation of the dual segmentation would be much more complex than in 2D to account for multi-boundaries as well as other non spherical topologies. Conversely, the algorithm presented in Appendix \ref{sec:algorithm} should be possible to extend to 3D.

\paragraph{}More robustness in quad extraction allows to relax constraints on the map generation algorithm. It is certainly a good news for the overall quad generation process, that could expect to be more robust, easier to implement and faster to run. In the volumic case, Hexex was a game changer for this reason, and pushing forward the robustness certainly increases the opportinuities to find applications for of quad/hex remeshing using global parameterizations. This work does not solve the problem but advance the theory in that direction.

\bibliographystyle{plain}  
\bibliography{refs}

\newpage
\appendix

\section{Algorithm}
\label{sec:algorithm}

As described in the core of the article, we do not have a robust algorithm yet, but an algorithm that competes with the state of the art using a simplified dual segmentation representation, and a heuristic to compute the sequence of operations OP1/OP2.

%Using the ideas of dual segmentation and its local modification operations OP1/OP2, we propose a new quad mesh extraction algorithm. A simplified dual segmentation representation is presented, then we explain how to construct it from a map $U$, how to simplify it by a sequence of OP1/OP2, and how to convert it to a quad mesh.

\subsection{Representation}

\begin{figure}[ht]
    \centering
    \includegraphics[width=0.7\linewidth]{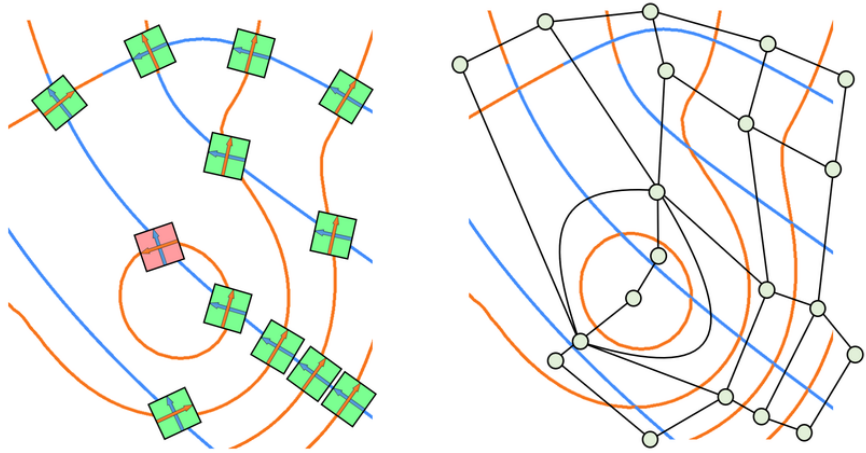}
    \caption{We have a graph of the network of iso, where nodes are located at intersection of iso-u and iso-v. Each node is represented by a square colored by it Jacobien sign (left), and where each edge associated to an iso curve and a direction. If we merge square edges linked by an iso curve, we obtain a quad mesh (right) represented with curved edges to show degeneracies.}
    \label{fig:datastruct}
\end{figure}

The dual segmentation is represented as a graph where edges are supported by iso curves, and where intersection of iso-$u$ and iso-$v$ are the nodes of the graph. Each node is represented by an isolated quad, having each edge orthogonal to the gradients (and opposite) of the map coordinates $\nabla u$, $\nabla v$, $-\nabla u$, or $-\nabla v$ (Figure \ref{fig:datastruct}). Nodes have integer map coordinates (they are located at the intersections of iso-$u$ and iso-$v$), and we know if they are reverted i.e. located in a folded part of the map. The set of quads is enriched by the "opposite" relation: a function that associate to each quad's edge, the quad's edge located at the other extremity of the graph's edge. It corresponds to the familliar "opposite" relation in the halfedge data structure: if quads are glued together by their opposite relation and vertices are merged accordingly, it produces the quad mesh.

\begin{figure}[ht]
    \centering
    \includegraphics[width=0.7\linewidth]{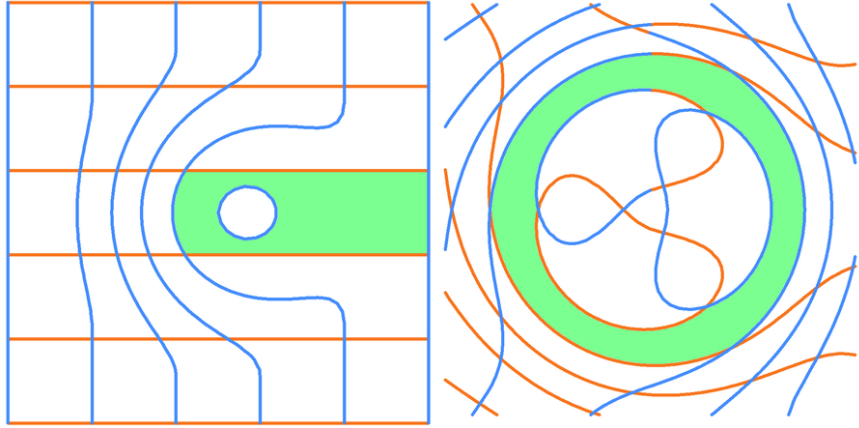}
    \caption{Single loop frontiers (left) and charts with multiple boundaries (green regions) are not representables in our data structure. }
    \label{fig:multiboundary}
\end{figure}

This structure is simple to work with, but has two limitations: an iso-curve that does not intersect another iso-curve will not be represented, and dual charts are assumed to be topological disks. This later property comes from the implicit representation of charts as sets of vertices to be merged in the final mesh according to the opposite relations. As a consequence, some configurations are not representable in our structure (Figure \ref{fig:multiboundary}).

\begin{figure}[ht]
    \centering
    \includegraphics[width=\linewidth]{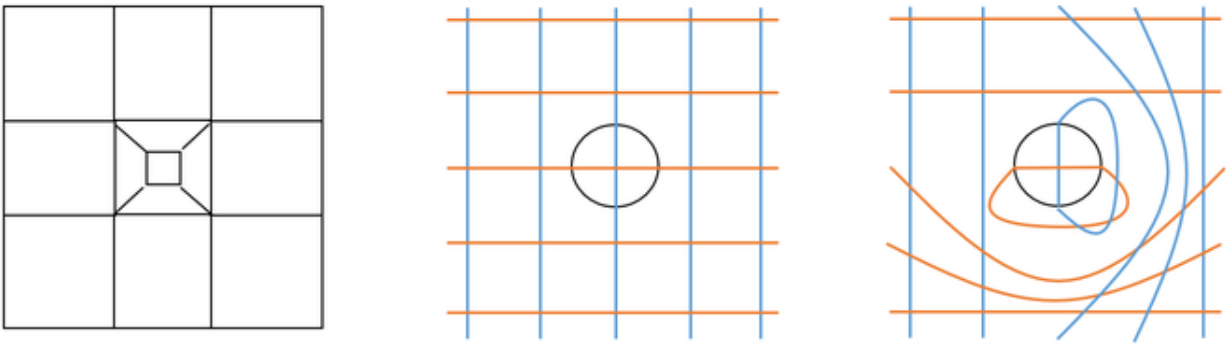}
    \caption{The quad mesh (left) corresponds to a dual segmentation (middle). Using one layer lift per dimension, we can produce a multi-boundary charts where ignoring one of the connex component would lose information about the original mesh structure.}
    \label{fig:failure}
\end{figure}

These limitations prevent our algorithm to have any proof of robustness, but are not that dramatic in practice. Our structure simply ignores boundaries having a single loop, and considers a chart with two boundaries as two different charts. In Figure \ref{fig:multiboundary}, the internal boundary will be ignored (left) or collapsed (right), and the expected result will be obtained. Conversely, we can construct a case of multi-boundary chart (Figure \ref{fig:failure}) where no boundary will be removed by the untangling process.

\subsection{Initialization}

The construction of our data structure requires to trace iso-u and iso-v on the surface. One could directly use the algorithm proposed in \cite{QEX} with exact arithmetic and end points that can be located on a triangle, and edge or a vertex of the triangle mesh.

However, iso of the dual are never forced by integer constraints to pass exactly through singularities or follow edges such as features or boundaries. As a consequence, there is no reason for a point located at iso intersections to lie on an edge or a vertex of the original mesh. 

We assume that all quads of our data structure are located inside a triangle to avoid deadling with many special cases: exact arithmetics, quads embedded in edges or vertices, with uncertain sign of the Jacobien, etc. Still, for this assumption to be reasonable, we sanitize the map before the segmentation to ensure curve tracing behaves correctly.

\subsubsection*{Sanitization}

The objective here is that all iso-u (resp. v) are defined as a set of segments whose endpoints are located on two edges of a triangle of the original mesh. Moreover, points located on both iso-u and iso-v must be located strictly inside a triangle.

To do this, we detect triangle edges that are located on an iso curve, or pass through a point with integer map coordinates. For these edges, we add a little perturbation (add a vector of norm $10^{-3}$ in a random direction) of the map coordinates of its extremities. This perturbation is propagated to other triangles corners sharing the same vertex, using the transition functions. Note that if both edge extremities are singularities, it is not possible to apply the perturbation, so we need to split the edge.   

Our implementation uses floating points, so failure examples can certainly be constructed by e.g. translate the map by $10^{30}$ so that there is not enough precision to represent the perturbation translation... In practice, we never encountered problems at this step.

\subsubsection*{Data structure initialization}

We first extract the quads: in each triangle of the mesh, an intersection of a iso-$u$ and iso-$v$ curve generates a quad. The quad is marked as \textbf{reverted} if $\det \nabla U < 0$ in that triangle.

We then connect them together: each quad has four edges, orthogonal to $\nabla u$, $\nabla v$, $-\nabla u$, or $-\nabla v$. For each quad edge: we trace the iso-curve in the direction given by the gradient of the other coordinate until we reach another quad. If two edges reach each other’s quads, we mark them as opposites. The iso-curve tracing algorithm has to cross transition functions as in \cite{QEX}, but is much simpler because there is a unique case at each step: crossing a triangle.

\subsection{Simplification}

The structure extracted in the previous section can be converted into a quad mesh, but this quad mesh is likely to be degenerated due to reverted quads and some missplaced index $-1$ due to saddle lifts. The first step of our algorithm will remove as many reverted quads as possible ({\bfseries untangling}), and the second step will perform some saddle lifts to lower the number of degenerated quads ({\bfseries index fairing}).

\subsubsection*{Untangling}
\label{sec:untangling}

\begin{figure}[ht]
    \centering
    \includegraphics[width=\linewidth]{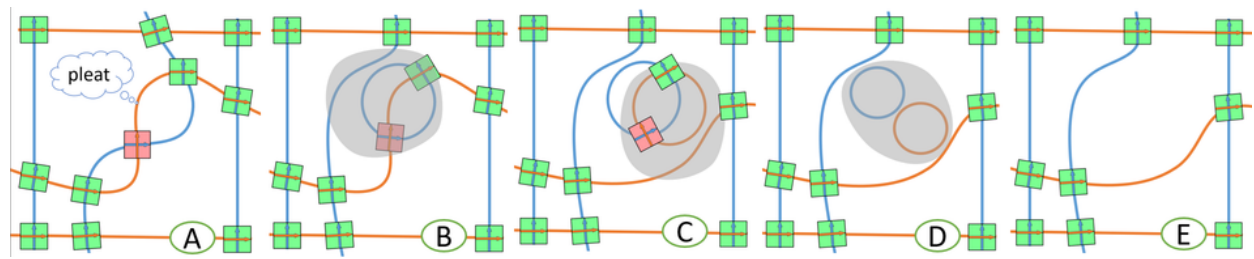}
    \caption{{\bfseries Pleat removal:} A pleat is detected as the opposite relation involving exactly one reverted quad (A), then it is deconnected from the rest of the graph by OP2 (B,C). Using OP1 disconnects u and v loops (D), that become isolated and removed (E).}
    \label{fig:untangle}
\end{figure}

The objective of untangling is to remove reverted quads. As we cannot remove them in the middle of reverted regions, we do it at the frontiers of reverted regions; we detect a pleat and remove the pair of quads involved in the pleat. This operation starts by isolating the pleat from the rest of the quads, before killing it, as illustrated in Figure \ref{fig:untangle}.  

The untangling step finds a pleat, iterates over each other edges connected the squares involved in the pleat and reconnects them to isolate and remove the pleat. The process continues until no more pleat can be found in the structure. At the end of the process, there is no more pleats, so each connected component of our structure is either fully reverted or has no reverted quads. In practice, there is always only one connected component, but we have no guaranty of this because of multi-boundaries charts that our structure cannot represent.

\subsubsection*{Index Fairing}
\label{sec:pairing}

\begin{figure}[ht]
    \centering
    \includegraphics[width=\linewidth]{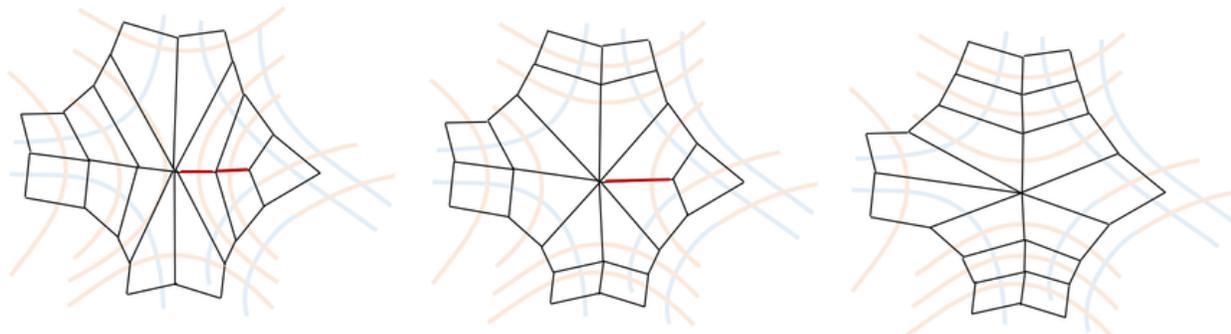}
    \caption{A path (red edges) is found to connect the valence 8 vertex to the valence 3 vertex (left). A first layer lift shorten the path (middle), and a second layer lift merges both singularities into a valence 7.}
    \label{fig:pairing}
\end{figure}

After the untangling step, we can extract a quad mesh that corresponds to a valid map $M$ as defined in \S \ref{sec:pb_settings}. However, it does not necessary corresponds to the expected combinatorial structure because of saddle lifts, and there is actually no way to determine which combinatorial structure is the best. At this point, all we can do to improve the quad mesh is to use heuristics.

In most cases, the saddle lift breaks a singularity of index $-\frac{1}{4}$ into two singularities of indices $\frac{3}{4}$ and $-1$. In the quads mesh, the index $\frac{3}{4}$ generates a vertex of valency $1$ that is obviously degenerated. It also happens with index $-\frac{1}{2}$ splitted into indices $\frac{1}{2}$ and $-1$, leading to a valence $2$ vertex in the quad mesh, that is geometrically degenerated for flat meshes, but may be acceptable for surfaces.

Our heuristic is to avoid configurations that will produce degenerated quads by local merges of singularities. The detection of singularities to be merged starts from a index $\frac{3}{4}$ and searches the shortest edge path to a singularity of index $\leq -\frac{3}{4}$. The merge is performed by a sequence of saddle lift that moves the singularity along this path (see Figure \ref{fig:pairing}).

The process finds all shortest paths starting from $\frac{3}{4}$ singularities, and simplifies the shortest is the length is under a threshold ($10$ edges in our case). The process is repeated until no more candidate is available. The same process is then performed for $\frac{1}{2}$ singularities.

\begin{figure}[ht]
    \centering
    \includegraphics[width=.6\linewidth]{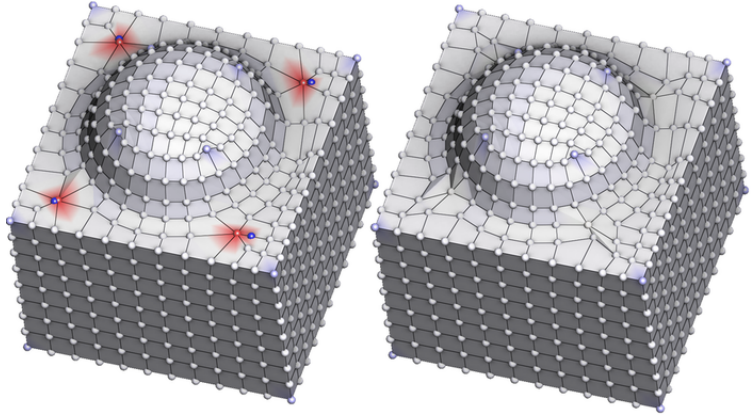}
    \caption{Quad meshes obtained without (left) and with (right) index fairing. Vertices color show their valence, to reveal degenerated quads.}
    \label{fig:indexfairing}
\end{figure}

In practice, can observe (Figure \ref{fig:indexfairing}) that index fairing plays a role even in quite simple cases. 

\subsection{Quad Mesh Extraction}

Our quad structure can be converted into a final quad mesh by identifying vertices such that the opposite-edge relationships match the classical opposite relation of halfedges in a mesh.

The combinatorial structure is obtained by merging vertices according to the opposite relationship of our data structure. The geometry is not clearly defined for maps with foldovers, but we know for each quad a point on the surface and its integer valued map coordinates. From these quad center points, we trace four curves on the surface, following diagonal directions in the map until a foldover or a point with integer + $(\frac{1}{2},\frac{1}{2})$ map coordinates is reach. If the map is valid, it founds exactly the good vertex position as in \cite{QEX}, ortherwise vertices are placed at the position that is the closest to the integer + $(\frac{1}{2},\frac{1}{2})$ map coordinates among all neigborg quads.

\subsection{Results}

We compare our algorithm with the author's implementations of \cite{QEX} and \cite{Hexex}. Hexex works only on tetrahedral meshes, so we extrude our triangulations before running Hexex. We use maps of various quality computed from existing datasets as well as maps designed to be challenging.

\paragraph{Dataset}

\begin{figure}[ht]
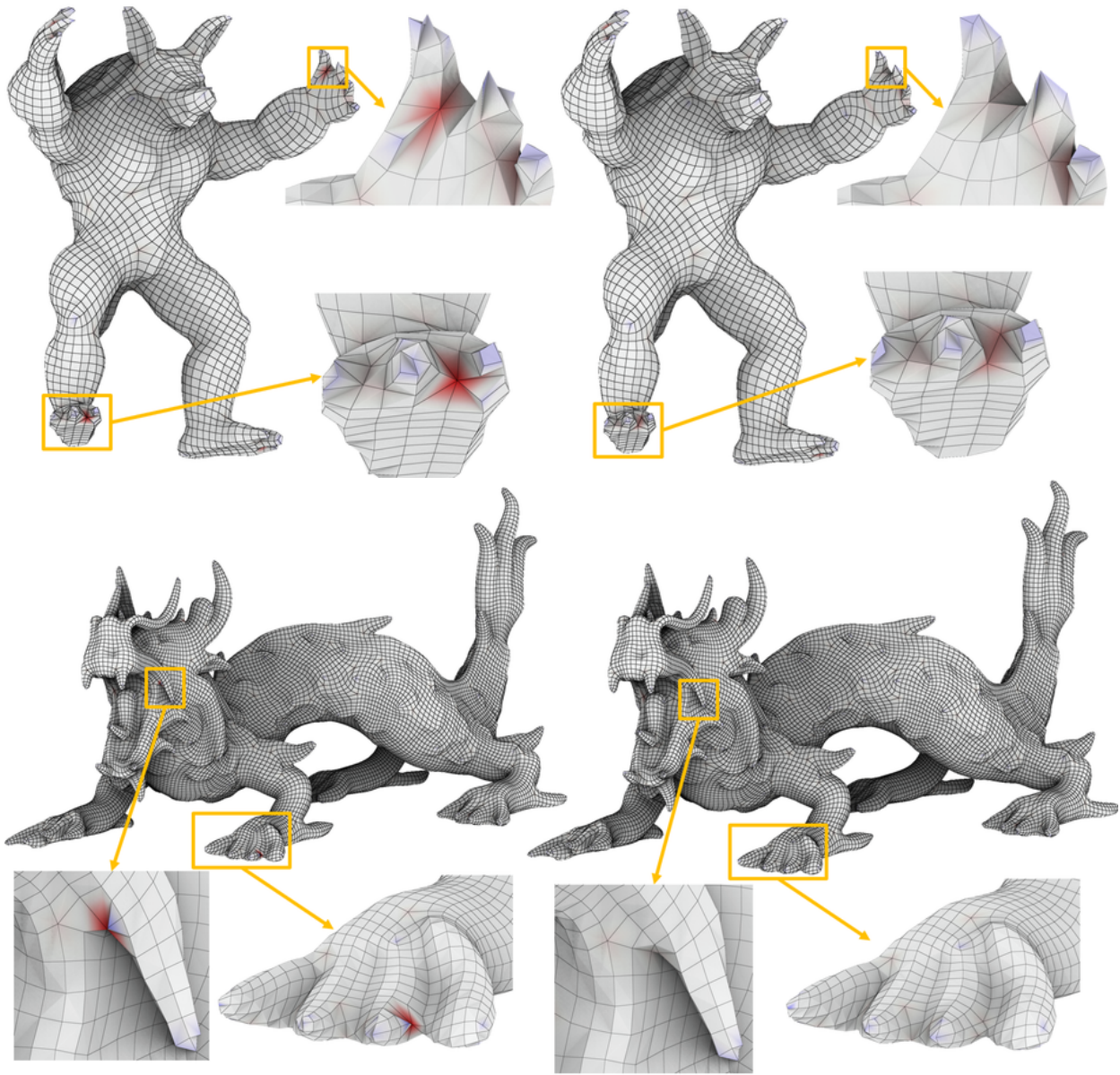

    \centering
    \includegraphics[width=\linewidth]{\FIGS{cmpqex3.png}}
    \includegraphics[width=\linewidth]{\FIGS{cmpqex2.png}}
    \caption{Using maps from Qex dataset, our algorithm (right) gives almost the same results than theirs (left). Our index fairing step reduces the number of low index singularities as illustrated in close-ups. Colors are used to show the valence of vertices.}
    \label{fig:cmpqex}
\end{figure}

\begin{figure}[ht]
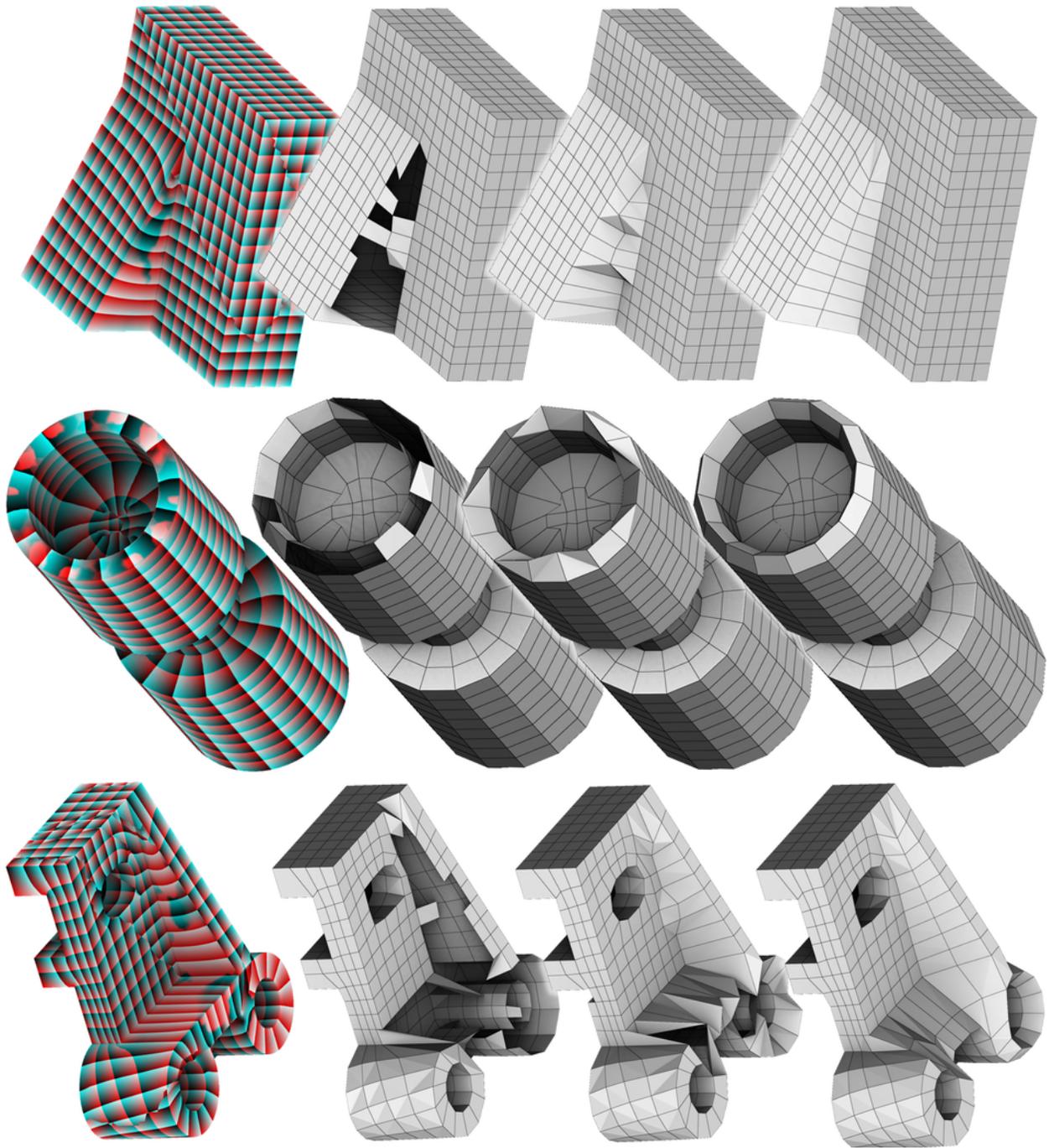

    \centering
    \includegraphics[width=\linewidth]{\FIGS{mambo1.png}}
    \includegraphics[width=\linewidth]{\FIGS{mambo2.png}}
    \includegraphics[width=\linewidth]{\FIGS{mambo3.png}}
    \caption{Quad meshes obtained by Qex, Hexex and our algorithm on CAD models. The map often merges features due to the frame field (upper row), the quantization (middle row) or both (lower row).}
    \label{fig:mambo}
\end{figure}

\begin{figure}[ht]
    \centering
    \includegraphics[width=\linewidth]{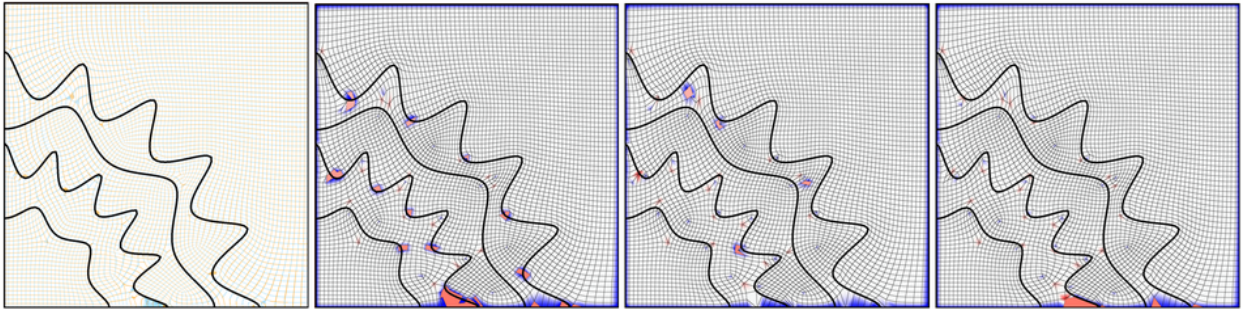}
    \caption{A folded map is computed on 2D domain with inner constraints (left). Qex and Hexex (middle left and right) generate meshes with holes, due to saddle lifts that have moved indices -1. Our algorithm (right) does not have the same issue.}
    \label{fig:cmpprevw}
\end{figure}

Maps provided by \cite{QEX} have good quality and our results are very similar to theirs (see \ref{fig:cmpqex}). We also generated lower quality grid preserving maps using field integration as geometric objective and rounding as quantization on the mambo dataset (Figure \ref{fig:mambo}) and some multi-material 2D flat models (Figure \ref{fig:cmpprevw}).

\paragraph{Toy challenges}

\begin{figure}[ht]
    \centering
    \includegraphics[width=.9\linewidth]{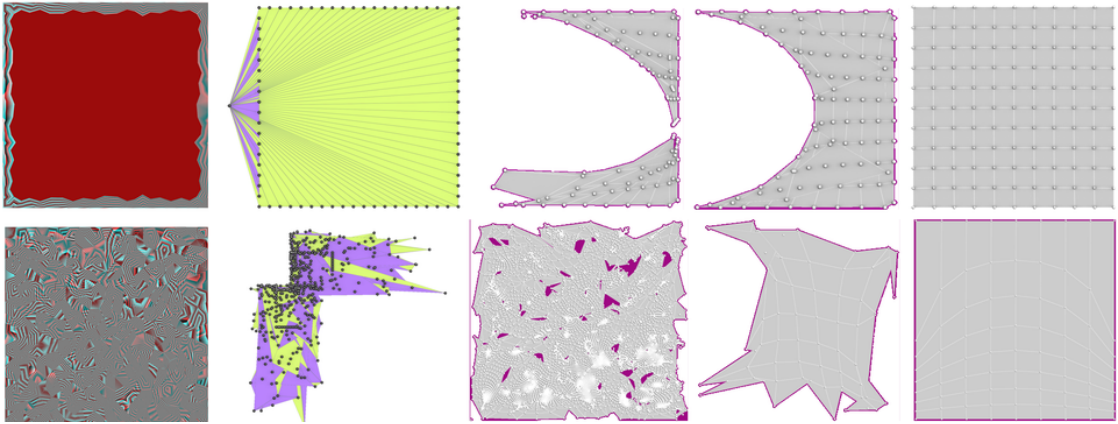}
    \caption{Stress tests; All map coordinates are set to (0,0) inside a square (upper row), and  a rotation for half the square plus a random noise are applied to all map coordinates (lower row). From left to right: combined map coordinates scalar fields, the map (reverted triangles are purple), Qex, Hexex and our results. {\bfseries Note:} Free boundaries are not supported by previous works, please don't mind the geometry.}
    \label{fig:stresstest}
\end{figure}

To evaluate the capacity of our algorithm to restore reverted regions, we start from a squared domain and take the identity as a map, then we move vertices with two strategies (Figure \ref{fig:stresstest}): placing all of them to the same point, and adding some transition functions and noise. Qex, Hexex and our algorithm are able to find the correct combinatorics in all cases, except Qex with the noisy map that is probably due their implementation (according to error messages).

\begin{figure}[ht]
    \centering
    \includegraphics[width=.9\linewidth]{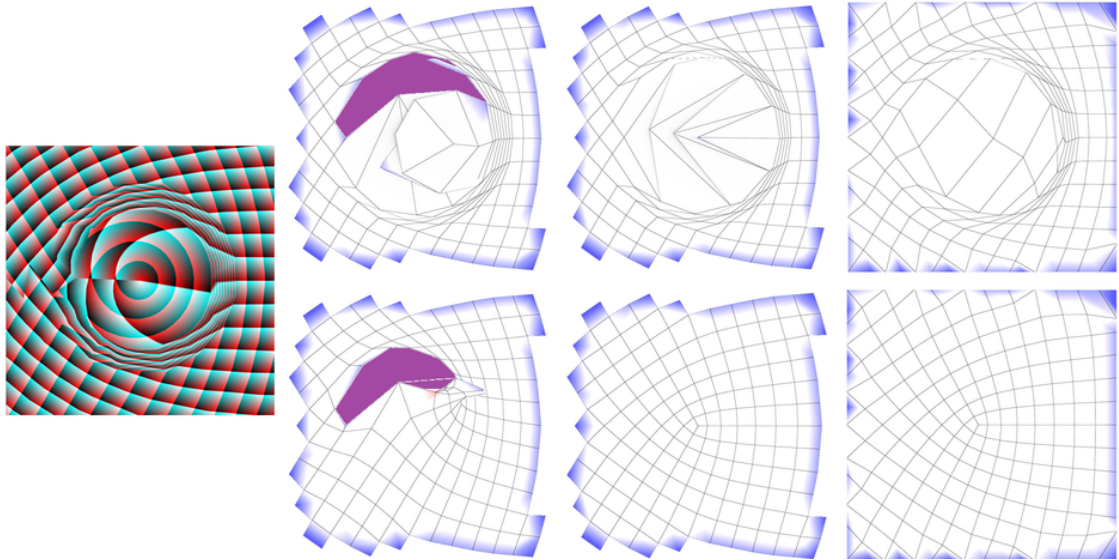}
    \caption{Singularity inside reverted region. From left to right, the map, Qex, Hexex and our algorithm. Lower row is smoothed for better visualization of the combinatorial structure.}
    \label{fig:reverted3}
\end{figure}

\begin{figure}[ht]
    \centering
    \includegraphics[width=.9\linewidth]{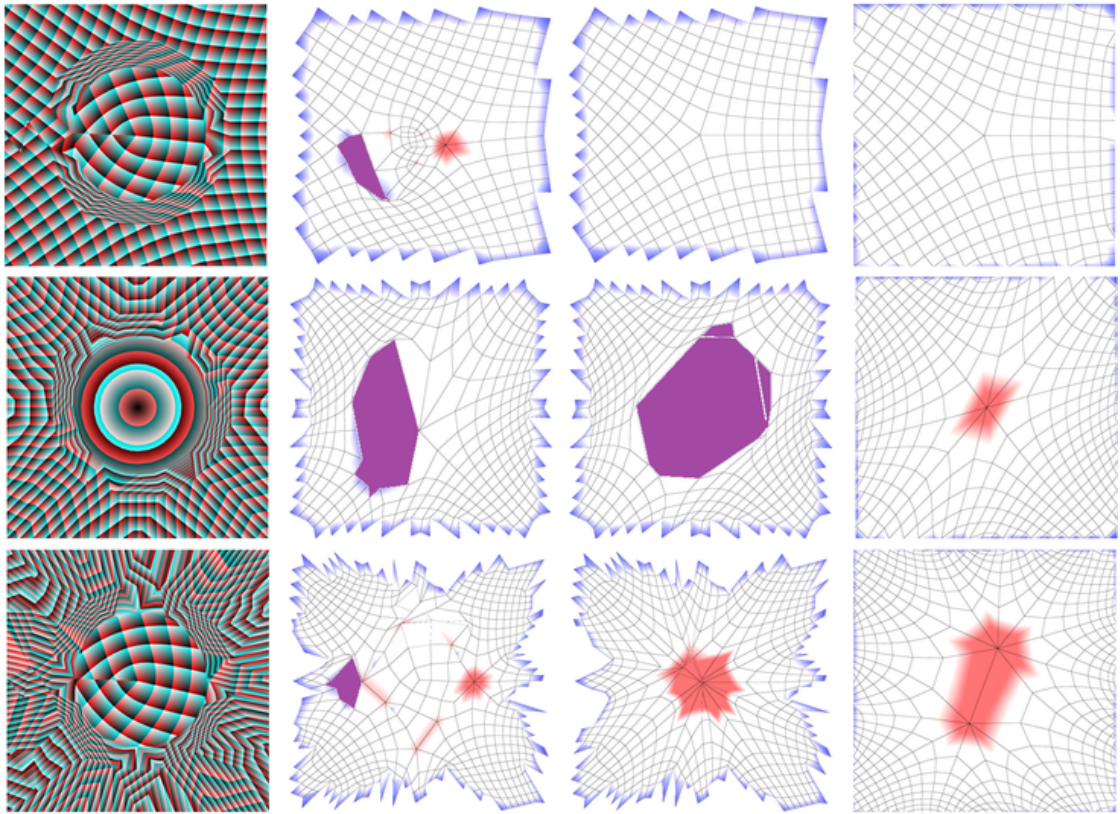}
    \caption{Same as in Figure \ref{fig:reverted3}, showing only smoothed meshes for other indices of singularity.}
    \label{fig:revertedother}
\end{figure}

A much more difficult challenge is to deal with foldovers interacting with singularities. Figure \ref{fig:reverted3} shows a first case where the a disk that constains a singularity of index $\frac{1}{4}$ is reverted. Hexex and our algorithm were able to recover the expected structure without producing a hole. Figure \ref{fig:revertedother} explore other cases with negative indices. Our results never have holes in the mesh, and we fail to obtain the initial configuration only with the index $\frac{-5}{4}$ singularity because there is no way to know that only one singularity was expected.


\begin{figure}[ht]
    \centering
    \includegraphics[width=.9\linewidth]{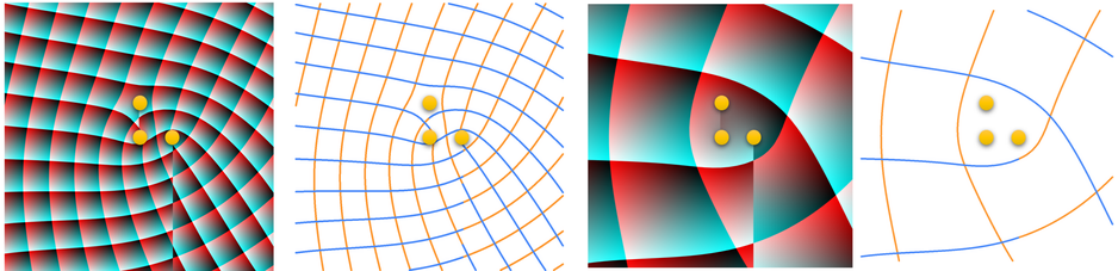}
    \caption{If singularities are separated in the map, the segmentation reflects it (left), otherwise that are merged in the same chart  (right).}
    \label{fig:multisingu}
\end{figure}

Another test was done with an invalid quantization (according to \cite{QGP}), and all algorithms (Qex, Hexex and ours) have merged the singularities that are not separated by the quantization in the map (Figure \ref{fig:multisingu}).

\section{Limitations and Discussion}
\label{sec:limitations}

\subsection{Representation of the dual segmentation}

We store intersections of iso-u and iso-v and link them by iso. Therefore we are not able to represent boundaries with only one adjacent chart, and charts with multiple boundaries. Actually, boundaries with only one adjacent chart are ignored, and charts with multiple boundaries are considered as multiples charts. As a consequence, the algorithm supports them, but with lost and corrupted informations.

We never observed any failure using our poor quality inputs, but we were able to constuct one in Figure \ref{fig:failure}. One can imagine scenari where such situations would be more frequent e.g. optimizing the map on a subset of the domain.

Beside the lost of robustness, ignoring some iso-value makes it impossible to track singularities. For exemple, if a valid map has a index -1 singularity and its map coordinates are locked, a layer lift would produce an iso curve around the singularity; the presence of this curve is a crucial information to go back to the original combinatorial structure. It does not matter for the current problem where we consider that the input is completely given by the map coordinates, but our structure would not be able to carry more information about singularities such as indices or map coordinates.

\subsection{Sequence of operations}

In our definition of the inputs, we assume that there exist a sequence of operation OP1 and OP2 that can transform our structure to the structure that would have been extracted from a valid (GP,det+) map. Our algorithm uses as heuristic a sequence that favor applying as many OP1 as possible to remove all reverted quads.

This strategy works very well in practice, but has no guaranty that multi-boundary charts will not be generated during the process. Moreover, if we were able to represent all charts configurations, this strategy would not be trivial to extend.

\subsection{Implementation details}

\paragraph{Geometry} The vertices position are not directly contained in the dual segmentation as it is in the primal. When the input map is valid, a simple post-processing step allows to get their positions from tracing in the map. When the map is not valid, both solution propose a fair solution for placing vertices, as there is no ground truth anyway.

\begin{figure}[ht]
    \centering
    \includegraphics[width=\linewidth]{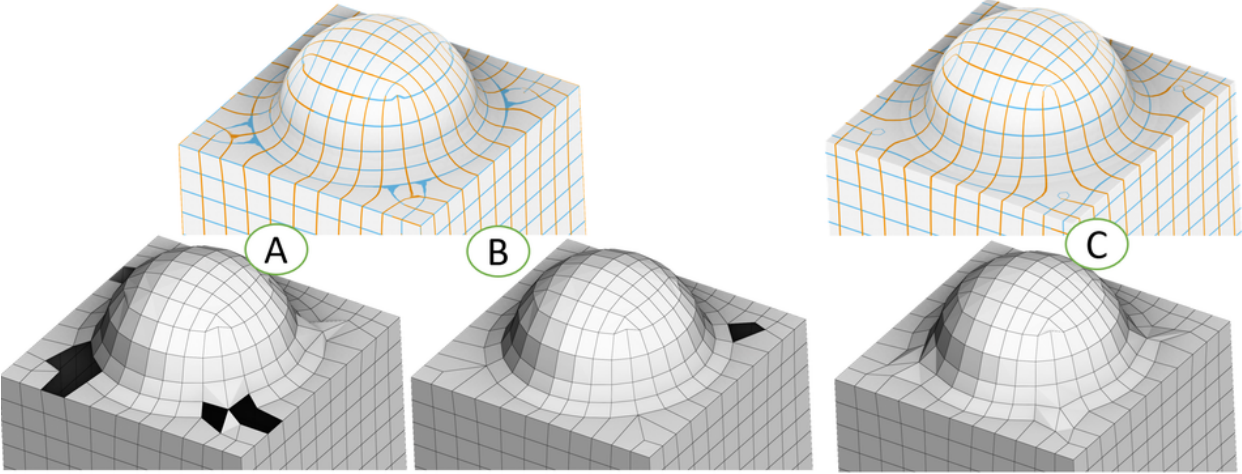}
    \caption{Our algorithm (C) based on the dual (upper right) on a simple CAD object with folded map. Qex (A) and Hexex (B), based on the primal (upper left) both fail to produce a closed mesh. However, our index fairing heuristic lost track of the feature curve, where Hexex preserved it.}
    \label{fig:cmpfeat}
\end{figure}

On the first hand, dual allows to better capture the boundary when the map is not aligned with it, but on the other had, the primal is more likely to fit the feature curves when they have been constrainted during the GP map optimization (Figure \ref{fig:cmpfeat}).

\paragraph{Numerical robustness}

Numerical accuracy is much less of an issue when tracing the dual than the primal, so we did not need to go further than floating points number representations and large epsilons during sanatization and geometric predicates for iso tracing.
Exact calculus and symbolic perturbations would certainly be requiered to proove the robustness of the process.

\paragraph{Index fairing heuristic}
Our heuristic manage to "repair" the usual configurations of saddle lifted nearby a singularity for improving the map distorsion. However, bad configurations may become even worse with our heuristic: for example, singularities that are not separated in the map may be merged into degenerated singularities (index $\geq 1/2$), and our heuristic will produce more mess when trying to pair it with a low index singularity.

\end{document}